\documentclass[twocolumn,eqnum,showpacs,prl,aps]{revtex4}
\usepackage{graphicx}

\begin{document}

\title{Impeded Growth of Magnetic Flux Bubbles \\ in the Intermediate State
Pattern of Type I
Superconductors}

\author{V. Jeudy, C. Gourdon}
\affiliation{
Groupe de Physique des Solides, Universit$\acute{e}$s~Paris~6~et~7,
CNRS~UMR~75-88\\
Tour 23 - 2, Place Jussieu - 75251 Paris cedex 05 - France}
\author{T. Okada}
\affiliation{Itoh laboratory, Division of Materials Physics, School of
Engineering Science, Osaka
University, 1-3 Machikaneyama-cho, Toyonaka-shi, Osaka 560-8531 - Japan}

\date{\today}

\begin{abstract}
Normal state bubble patterns in Type I superconducting Indium and Lead slabs
are studied by the
high resolution magneto-optical imaging technique. The size of bubbles is
found to be almost
independent of the long-range interaction between the normal state domains.
Under bubble diameter
and slab thickness proper scaling, the results gather onto a single master
curve. On this basis, in the
framework of the ``current-loop'' model [R.E. Goldstein, D.P. Jackson and A.T.
Dorsey,  Phys. Rev. Lett.
{\bf76}, 3818 (1996)], we calculate the equilibrium diameter of an isolated
bubble resulting from the
competition between the Biot-and-Savart interaction of the Meissner current
encircling the bubble and the
superconductor-normal interface  energy. A good quantitative agreement with the
master curve is found over two decades of the magnetic Bond number. The
isolation of each bubble
in the superconducting matrix and the existence of a positive interface
energy are shown to preclude any
continuous size variation  of the bubbles after their formation, contrary
to the prediction of
mean-field models.
\end{abstract}

\pacs{05.65.+b, 74.25.Ha}

\maketitle

A great variety of quasi-two-dimensional, biphasic systems present a spontaneous
formation of domain patterns:
magnetic  liquids \cite{elias}, Langmuir monolayers \cite{mcconnell},
sub-monolayer of adsorbed atoms
\cite{plass}, ferro- and ferrimagnetic films \cite{seul},
intermediate state (IS)
in Type-I superconducting (SC) materials \cite{huebener1}...
These structures are mostly interpreted as resulting from the balance
between long-range repulsive, electrostatic, magnetic or elastic
interactions between domains and
short-range attractive interaction associated with a positive interface
energy. The observed patterns are generally disordered and consist of bubbles and of branched and intricate fingered structures (lamellae). At present the mechanisms of the formation of these structures are theoretically actively studied  \cite{langer,goldstein,cebers}. In particular, for magnetic fluids, the instabilities of bubble circular shape was shown to produce fingered structures which are similar to
those observed experimentally \cite{langer}. The same mechanism was proposed for the IS in Type I superconductors \cite{dorsey}. However little is known even about the static properties of bubble patterns \cite{note0}. This question is of prime importance for the study of IS patterns formation since normal state (NS) bubbles form the early stage of the IS when the magnetic flux starts to penetrate into SC samples \cite{huebener1}.

IS patterns are observed in slabs placed in a  perpendicular magnetic field. They consist of SC and NS, flux-bearing
domains \cite{huebener1}.
Former studies were essentially focused on the lamella structures \cite{huebener1}.
The free energy of a one-dimensional lattice of infinitely long and parallel stripes was first calculated by Landau
\cite{landau}. The field-dependent predicted and measured periods of the stripes were
found in good agreement
\cite{miller}. Subsequently, their comparison
became a conventional method for determining
the interface energy of Type-I SC materials. The formation of lamellae was recently
re-examined by Goldstein, Jackson and
Dorsey in the framework of a ``current-loop''
model \cite{dorsey}. These authors propose to consider IS patterns as a set of
domains of arbitrary shapes with
vertical domain walls and bounded by current loops interacting in the free
space above and below the slabs
\cite{dorsey}. When applied to the stripe pattern the model predicts
equilibrium periods close to those found
using the Landau model, thus indicating that both models essentially capture the same physics. As the model is formulated for
arbitrary domain shape, it opens the way to study the formation of bubble patterns whose conditions of existence and control parameters are not well understood. To our knowledge, the only calculation
of the free energy of an hexagonal lattice of bubbles uses an approximate
expression of the magnetic energy
interaction
\cite{goren}. Subsequent experiments found a field-dependent bubble spacing
different from the predicted one. They also yield a smaller interfacial tension than the one deduced
from the studies of stripe patterns
\cite{goren,huebener,farrell,huebener2}. In view of these scarce and contradictory results, it cannot be established whether bubble patterns correspond to a quasi-ground state as it is the case for stripe patterns.  Futhermore, the onset of the formation of the IS was shown to result from the penetration of bubbles from the edges of samples \cite{chimenti}. The magnetic flux penetration is controlled by an energy barrier of geometrical nature \cite{jeudy,castro}. This raises the question of the respective contributions of the mechanism of flux penetration and of the balance between long-range and short-range interactions on the formation of bubble patterns.

This letter presents a systematic study of NS bubble patterns as
a function of the
SC  material, the slab thickness
and the applied magnetic
field. Contrary to the lamellae width, the diameter of the bubbles is found to be independent of the mutual interaction between flux-bearing domains. We discuss the origin of these different behaviors in terms of magnetic flux penetration mechanisms.


The domain patterns are observed with the high resolution Faraday
microscopy technique which probes the normal component of the local induction at
the top surface of a superconductor. Experimental details are given
elsewhere \cite{gourdon}.
The SC Pb slabs were cut out from
GoodFellow
$99.9
\%$ pure and annealed
$25$ and $120 \pm 1$ $\mu m$ thick foils. The magneto-optic layer (MOL)
consisted of a $1500$ {\it \AA} EuS film  evaporated on a glass substrate and covered with a $600$ {\it \AA} Al mirror. The Pb slab
was compressed against the mirror. The In slabs ($0.6, 1.1, 1.5, 2.2, 10.0
\pm 0.1$ $\mu m$
thick) were obtained by evaporation directly onto MOLs. The MOLs consisted of
CdMnTe/CdMgTe semiconductor heterostructures grown by molecular beam epitaxy
\cite{gourdon2}. The samples were immersed into superfluid helium
at temperatures $T\leq 2$ $K$. They were subjected to
an increasing
perpendicular magnetic field $H$ whose maximum value equals 60 $mT$.

\begin{figure}[h]
\includegraphics[width=0.8\linewidth]{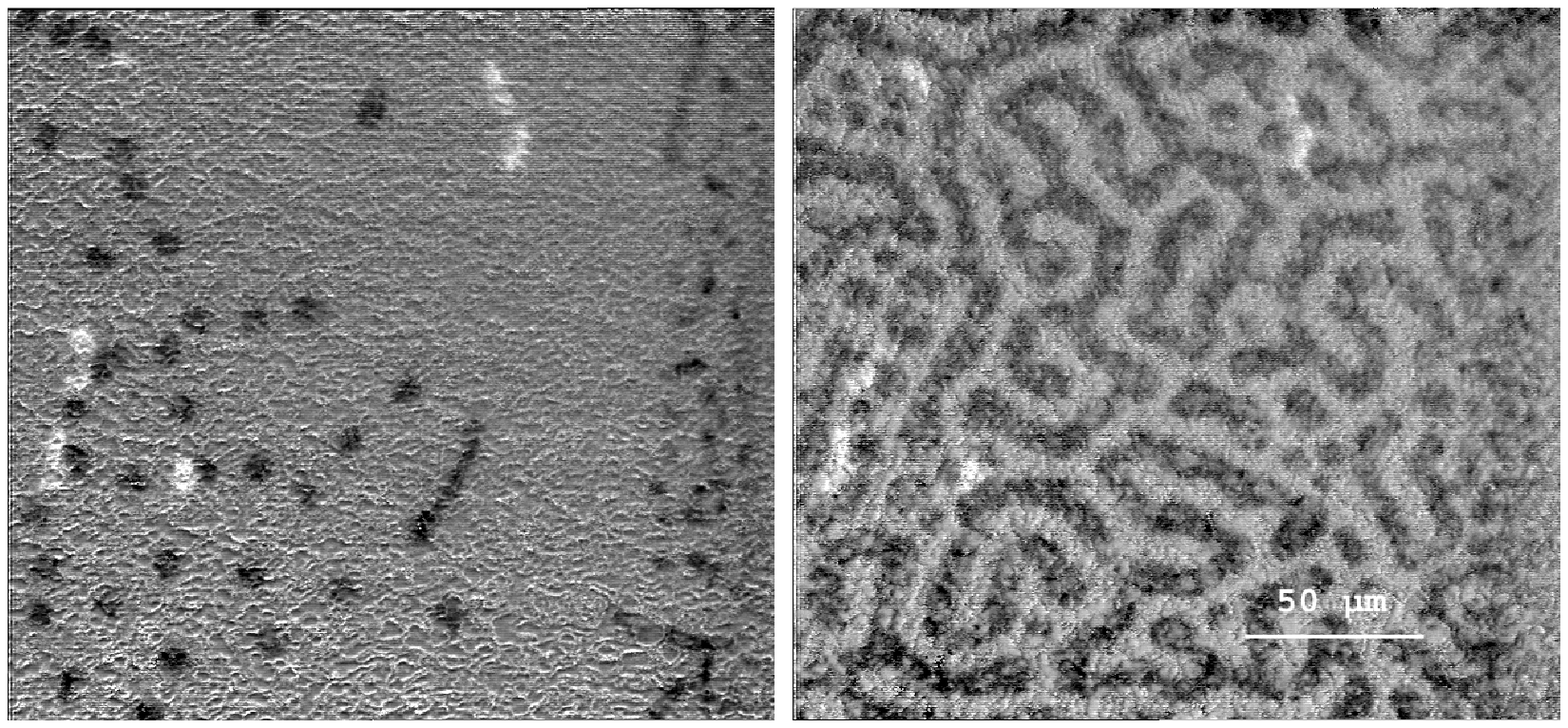}
\caption{Intermediate state pattern on the edge of the 10 $\mu m$ thick
superconducting Indium slab for $h=0.07$ (left) and $h=0.41$ (right) \textit{at} $T=$ 2
$K$. The edge of the In slab is along the right edge of the images. Normal and superconducting domains appear in black and gray,
respectively. The few white domains correspond to magnetic flux which was trapped at $h=0$ (details on image processing are given in Ref.  \cite{gourdon}).}
\label{image}
\end{figure}

Fig. \ref {image} shows typical IS patterns observed on the edge
of a 10
$\mu m$ thick Indium slab for two values of the reduced applied magnetic
field $h=H/H_c$, where $H_c$ is the thermodynamical critical field. Increasing $h$ results in the penetration of the
magnetic flux from the edges
of the slab which is revealed by a significant increase of the density of NS
domains. At low
$h$-value (left image), NS domains essentially consist of almost circular
bubbles. They were systematically observed over a
limited range of low $h$-values. At higher $h$-value (right image), lamellae have appeared. They progressively form labyrinthine structures.
%
%
%
While most of the lamellae are connected to the edges of
the slab from which the
magnetic flux enters, bubbles are isolated in
the SC matrix
and separated from the edges by a full diamagnetic band ($\approx 50 \mu m$ large). Let us note firstly that the interaction
between isolated NS
domains is repulsive and secondly that the diamagnetic band reflect the
presence of the geometrical energy barrier which prevent spontaneous flux
penetration on the edges \cite{jeudy,castro}. Therefore different formation
and growth
mechanisms are expected for bubbles and lamellae.


In order to get more insight into this question, the variation of the bubble diameter $2R$
and the lamella width $W$
were measured systematically as a function of $h$. They were then compared
to their respective equilibrium values $2R_{eq}$ and 
$W_{eq}$, calculated for regular arrays. Fig. 2. presents
the results obtained for a  10
$\mu m$ thick In slab (left) and for a 120 $\mu m$ thick Pb slab
(right). For the lamella pattern,
$W_{eq}$ is calculated from Eq.\  (3.23) and (4.8) of Ref. \cite{dorsey}.
For the bubble pattern, $R_{eq}$ is calculated in the framework of the
``current-loop'' model
\cite{dorsey}. NS bubbles with radii $R$ are assumed to be arranged in an
hexagonal lattice of period $a$ in
an infinite  slab of thickness $d$. The magnetic field in
the bubbles is equal
$H_c$ \cite{dorsey}. From the constraint of global flux conservation the area
fraction of NS domains $\rho_n =2\pi R^2/\sqrt{3}a^2$ is equal to $h=H/H_c$.
The interface energy
$E_{int}$ is the product of the interfacial tension
$\sigma_{ns}=(H_{c}^{2}/8 \pi)\Delta$ by the total area of the interfaces
$2N\pi Rd$ where $N$ and $\Delta$ are the total number of bubbles and the
``wall'' energy parameter, respectively. The
magnetic energy per unit area resulting from the self and mutual interaction between the
screening currents flowing at the
interfaces is found equal to:

%
\begin{equation}
\label{energylattice}
\epsilon_m =  - \frac{y^2N_b}{3
\pi}\sum_{m=-\infty}^{+\infty}\sum_{t=-\infty}^{+\infty}\frac{J_1^2\left({ys}
\right)}{s^2}\left({1-\frac{1-\exp{(-\frac{2\pi ds}{a})}}{\frac{2\pi
d s}{a}}}\right)\ ,
\end{equation}
%
%
with $s=\sqrt{4(m^2+t^2-mt)/3}$, $y=\sqrt{2\pi\sqrt{3}\rho_n}$. $J_1$ is
the Bessel function of
the first kind, $N_b=d/\pi\Delta$ is the magnetic Bond number \cite{dorsey} and $\epsilon_m$ is expressed in units of $\sigma_{ns}$.
The equilibrium period $a_{eq}$ is obtained by minimizing the
reduced total energy per unit area
$\epsilon(a)=\epsilon_{int}+\epsilon_{m}$. $R_{eq}$
is then obtained from flux conservation as
$R_{eq}=a_{eq}\sqrt{\sqrt{3}h/2\pi}$.
$H_{c}(T)$ was
assumed to follow a Bardeen-Cooper-Schrieffer temperature variation:
$H_{c}(T)=H_{c}(0)(1-(T^2/T_c^2))$.
$H_{c}(0)$ is 28.2 $mT$ and 80.3 $mT$ and $T_c$ is 3.4 $K$ and 7.2 $K$
for Indium and Lead,
respectively. $\Delta (T)$ was assumed to follow the empirical law $\Delta (T) =
\Delta (0) /\sqrt{1-(T/T_c)}$ as proposed in Ref. \cite{sharvin}. $\Delta (0)$-values were
taken from the literature: for Pb,
$\Delta (0) = 0.056\: \mu m$\cite{huebener1} ; for In, $\Delta (0) =
0.33\: \mu m$
\cite{sharvin}.

%
\begin{figure}[htpb]
\includegraphics[width=0.9 \linewidth]{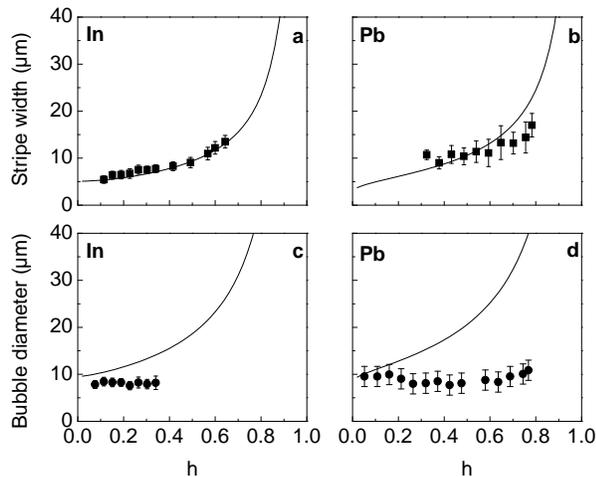}
\caption{Mean lamella width $W$ (top) and bubble diameter $2R$
(bottom) as a function of the reduced field
$h=H/H_c$ for a 10 $\mu m$
thick In slab (left) and a 120 $\mu m$ thick Pb slab (right). The error bars represent the full width at half maximum of the distributions of lamella width or bubble diameter. The solid lines are the equilibrium values $2R_{eq}$ and $W_{eq}$.}
\label{2R-versus-h}
\end{figure}
%

For the lamellae, $W$ and $W_{eq}$ present a good quantitative agreement, as
often reported in the literature (see Fig. 2a and 2b). The
slight discrepancy obtained for Pb, when $h>0.65$ may be attributed to the fact
that an important fraction of the lamellae remains isolated among bubbles.
Surprisingly, the measured bubble diameter $2R$ is
found to remain almost constant, as $h$ is increased, in disagreement with the theoretical
predictions (see Fig. 2c and 2d). The maximum ratio between $2R_{eq}$ and $2R$ is of the order
of 1.6 and 4 for In and Pb,
respectively. However, $2R$ approaches $2R_{eq}$
for $h \rightarrow 0$, thus suggesting that the
disagreement does not originate from the approximations used in the model
of Ref. \cite{dorsey}.

\begin{figure}[h]
\centering \includegraphics[width=0.75\linewidth]{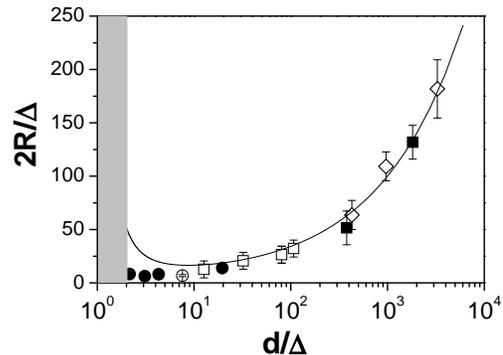}
\caption{Semi-log plot of the reduced bubble diameter $2R/ \Delta$ versus
reduced
sample thickness $d/\Delta$. The filled circles and squares were obtained with
In and Pb slabs, respectively. The empty squares, lozenges and circles are
reported from Ref.
\cite{huebener2,huebener,farrell}, for Pb, Hg and In, respectively. The gray region corresponds to Type-II superconductivity for very thin slabs. The solid curve is the equilibrium
diameter (Eq. \ref {eqradius}).}
\label{2R-versus-d}
\end{figure}

To clarify this point, the bubble diameter $2R$ was measured for different
slab thicknesses $d$ and compared to $2R_{eq}^0$, the limit of
$2R_{eq}$ calculated for $h
\rightarrow 0$. In this limit, the magnetic energy is only determined by the self interaction of the screening currents flowing at each bubble interface. We find that $2R_{eq}^0$ is the solution of the implicit
equation :
%
%
\begin{equation}
\label{eqradius}
         N_{b}= \frac{3(1-k^{2})}{k^2} \left[ 1+ \frac{1}{k^3}((k^{2}-2)
E(k)+2(1-k^{2})
K(k)) \right]^{-1}.
\end{equation}
%
%
with $k^{2}=4 {R_{eq}^0}^2 /(d^{2} + 4 {R_{eq}^0}^2)$. K and E are the
complete elliptic
integrals of the first
and the second kind, respectively. Eq. \ref{eqradius} is transformed into a relation
between the reduced variables $2R_{eq}^0/\Delta$ and $d/\Delta$ as plotted
in Fig. \ref{2R-versus-d}.
The same figure reports the results
obtained by us and by other authors with different
SC materials (In, Pb, Hg) \cite{hg}. Scaling
$2R$ and
$d$ by the wall energy parameter
$\Delta$ allows to gather all the measured diameters onto a single
master curve. This
demonstrates that
$2R/\Delta$ and
$d/\Delta$ are appropriate reduced variables to describe the bubble patterns.
Furthermore, the magnetic Bond number $N_b=d/\pi\Delta$ varies over the full range
of existence of non-branching IS
patterns ($1<N_b<1000$)\cite{goren,hubert}.For smaller Bond numbers the gray region in Fig. \ref{2R-versus-d} indicates the occurrence of Type-II superconductivity below a critical thickness $d_c$. Indeed, no IS domains were observed for the
thinnest In slab ($d=0.6 \mu m$). 
%
%
%
%

The comparison between the master curve and the prediction of
Eq.
\ref{eqradius} shows that two ranges of
$d/\Delta$-values can be distinguished.
For
$d/\Delta <30$, $2R$-values are found slightly smaller than $2R_{eq}$. In this range of thickness, the assumption of a constant $\Delta$ is no longer valid. As the thickness decreases towards the critical value $d_c$ the interfacial tension (positive for a Type-I superconductor and negative for a Type-II one) should decrease. Indeed adjusting the experimental data to the
predicted curve leads to a continuous decrease of $\Delta$
 with decreasing $d$ (not shown here). The critical thickness at which
$\Delta \rightarrow 0$ was found equal to $d_c=0.8\; \mu
m$, a value consistent with
$d_c \approx 0.9-1\; \mu m$  reported for In in Ref. \cite{goren}.
Therefore, the poor agreement found for
$d/\Delta <30$ most likely originates from the reduction of the
interface energy when the slab thickness
is decreased.
For  $d/\Delta >30$, the master curve presents a very good quantitative agreement with the prediction of Eq.
\ref{eqradius}.
This shows that the bubble mean diameter is determined by the balance between the interfacial tension and the  \textit{self}-interaction of the screening current flowing at  the bubble interface. While the bubble diameter remains constant when $h$ increases, as shown in Fig.
\ref{2R-versus-h}, the mutual interaction between the bubbles serves to adjust the mean distance between them so that the area density of NS domains $\rho_n$ is very close to the equilibrium value. We calculate that the bubbles free energy is only $\approx$ 1 \% larger than the equilibrium value. The reason is that volume terms depending on $\rho_n$, but not on the period, are dominant in the free energy. Hence the bubble system is only in a very slightly out-of-equilibrium state.

These results raise the question of the growth mechanisms of  NS domains. For the lamellae, the good agreement between the predicted and the measured width suggests that their growth is continuous and reversible, as assumed by the IS models. This most likely results from the fact that lamellae are connected to the edges of the slab, thus allowing infinitesimal amounts of magnetic flux to penetrate continuously from the exterior. This is not the case when NS domains are isolated within the SC matrix.
 As the flux density is uniformly equal to zero in the matrix, it follows from the constraint of flux conservation that the increase of the size of an isolated NS bubble has to result
from the incoming of another NS bubble crossing the
surrounding SC region. The fusion of these two bubbles is impeded by their repulsive interaction. Moreover, surface tension prevents the
formation of bubbles of size much smaller than
$2R_{eq}^0$.
This precludes the continuous and reversible growth of bubbles. Therefore they have to keep the size acquired during their formation as it is observed experimentally. A similar phenomenon should be encountered in other physical systems for which the mechanism of growth of isolated domains requires the migration of particles or of flux lines through a second phase. For example, in a ferrofluid confined in a Hele-Shaw cell with an immiscible non-magnetic liquid, the inhibited migration of magnetic particles between domains should prevent their size variation.

The early stage of bubble formation and migration is not accessible experimentally. Bubble velocities close to the sample edges ($> 1\mu m/\mu s$)  \cite{chimenti} are beyond our experimentally measurable velocities ($\approx 1\mu m/ s$). Therefore only a qualitative and partial understanding of bubble penetration can be inferred
from the results presented above. The concentration of bubbles was observed to increase with $h$ essentially
while the
diamagnetic band is present (see Fig. \ref{image}). The
bubbles should therefore be formed in the
region of the edges and cross the diamagnetic band to reach the sample
interior. The characteristic
sizes of the domains observed on the edges are smaller than those observed
within the bulk. This
suggests that the bubbles have to grow and to come unfastened from the IS structures present on the edges of the slab. The motion of the bubbles towards the center of the slab is driven by the magnetic interaction between the flux bearing domains and the magnetic field around the slab. However, the size of the bubbles was found to be independent of the aspect ratio of the slabs. This indicates that, even if the magnetic field gradient plays a role in the instability giving birth to a bubble, its size is essentially controlled by the competition between the surface tension and the self-interaction of the screening current. It would be of great interest to determine to what extent the size of the domains observed in other diphasic systems results more from specific mechanisms of the formation of domains than from the competition between long and short-range interactions.

Experimental evidence of the branching instabilities of circular NS bubbles, predicted in Ref.
\cite{goldstein}, was not found. Branched domains always bear a much
larger magnetic flux than bubbles. As a result the branching instabilities of bubbles are unlikely to be the prevalent mechanism for the formation of fingered structures. Whether, as in the case of bubbles, this formation results from an instability of the IS structure located on the edge of the sample remains to be investigated.

\begin{acknowledgments}
The authors would like to thank A.O. Cebers for numerous and fruitful
discussions.
\end{acknowledgments}


\begin{thebibliography}{breitestes Label}
\bibitem{note0} Instead of the terms
``flux tubes'' or ``flux spots'' found in the literature dedicated to IS
studies, we here adopt the conventional designation used for pattern studies.
\bibitem {hg} For Hg we take $\Delta (0)$ equal to $0.084\: \mu m$,
i.e. three times smaller than the value proposed in Ref. \cite{farrell}. The overestimation of $\Delta (0)$ in Ref. \cite{farrell} results from the use of an approximate model to calculate the magnetic energy.
\end{thebibliography}
\end{document}